\documentstyle[12pt,aaspp4]{article}

\begin{document}

\title{Temperature and Emission-Measure 
Profiles Along Long-Lived Solar Coronal Loops 
Observed with TRACE}

\author{Dawn D. Lenz,\altaffilmark{1} Edward E. DeLuca,\altaffilmark{2}
       Leon Golub,\altaffilmark{2} Robert Rosner,\altaffilmark{1}
       and Jay A. Bookbinder\altaffilmark{2}}

\altaffiltext{1}{Department of Astronomy and Astrophysics,
       University of Chicago, 5640 South Ellis Avenue, 
       Chicago, IL 60637; lenz@oddjob.uchicago.edu, 
	rrosner@oddjob.uchicago.edu}

\altaffiltext{2}{Smithsonian Astrophysical Observatory, 60 Garden Street, 
Cambridge, MA 02138; edeluca@cfa.harvard.edu, lgolub@cfa.harvard.edu, 
jbookbinder@cfa.harvard.edu}

\author{Received 1999 February 22; accepted 1999 March 29}

\author{To appear in the Astrophysical Journal Letters}

\begin{abstract}

We report an initial study of temperature and emission measure 
distributions along four steady
loops observed with the Transition Region and Coronal Explorer (TRACE) 
at the limb of the Sun.  The temperature diagnostic is the filter
ratio of the extreme-ultraviolet 171 \AA\ and 195 \AA\ passbands.  The
emission measure diagnostic is the count rate in the 171 \AA\ passband.  
We find essentially no temperature variation along the loops.  We compare
the observed loop structure with theoretical isothermal and 
nonisothermal static loop structure.

\end{abstract}


\section{Introduction}

The Transition Region and Coronal Explorer (TRACE) is producing
a wealth of high-quality, high-cadence, high-resolution data for the
solar corona in the extreme ultraviolet (\cite{sta96}; \cite{wbj97}; \cite{h99}) 
that allow us to
probe the spatial and temporal structure of the corona in
unprecedented detail.  

The detailed properties of the corona are central to solving the
coronal heating puzzle.  Coronal loops are the most basic coronal
structure, as evidenced by Yohkoh and most recently by TRACE.
Theoretical studies of coronal structure and heating have thus
focused on understanding loops (\cite{lm75}; 
\cite{vr78}; \cite{spv81}; \cite{rf82}; 
\cite{j92}; \cite{cps93}).  
Previous and current EUV studies have
found isothermal, hydrostatic loop structure (\cite{gj75}; Aschwanden 
et al. 1999a, 1999b),  
while broadband X-ray analysis and
theoretical calculations have suggested that coronal loops have temperature
maxima at their tops (\cite{rtv78}; \cite{spv81}; \cite{kt96}).  

Here we report a first look at temperature and emission-measure 
structure along the
axes of coronal loops observed with TRACE (Fig. 1).  The temperature
diagnostic we use is the 171 \AA\ /195 \AA\ filter ratio (Fig. 2a).
The emission measure (EM) diagnostic is the 171 \AA\ passband 
count rate (DN/s/pixel, DN$=$data number) (Fig. 2b), as calculated using
the CHIANTI atomic database (\cite{dlm97}).

\section{Loop Observations}

The data consist of observations in the 171 \AA\ (Fe IX) and 195 \AA\
(Fe XII) TRACE filters of four loop systems observed at the limb of the Sun
(Fig. 1).  The instrument resolution is 1\farcs0.   
We attempted to choose relatively isolated loops 
that extend above the limb of the Sun in order 
to minimize non-loop background flux and 
projection effects.
Since we focus here on steady loop structures, the loops selected for
measurement showed little morphological variation during the selected 
$1 - 2$ hour
intervals (the loops were usually observed at high cadence 
for considerably longer
times, up to six hours, but we restrict the time interval in an
attempt to minimize effects of morphological evolution and 
of rotation onto and off of the limb).
The data set for each loop thus consists of approximately one hour
of high-cadence observations, each $10 - 40$ seconds long.

To investigate the variation of temperature along the loop, we then
selected four subimages of each loop representing (1) an area near the
base of the loop (roughly 1/5 of the distance to the loop top), 
(2) an area approximately 1/3 of the distance to
the loop top, (3) an area approximately 2/3 of the distance to the
loop top, and (4) an area roughly at the loop top.  Rather than
consider the entire loop length, we consider the half of each
loop that shows the least overlap with adjacent structures.
The subimage 
selection attempted to contain an adequate number of pixels from the
loop of interest to spatially average, 
while excluding pixels from the background and/or
adjacent coronal structure.  Each subimage contains
a few hundred to a few thousand pixels.

Precise loop length determinations would   
require analysis of possible projection effects; however, approximate loop
lengths can be inferred from Figure 1.  
Rough estimates of the loop semilengths are $L = 10^{10}$ cm for 
loops (a), (b), and (d) and $L = 5 \times 10^{9}$ cm for loop (c).

\subsection{Data Reduction}

The data were first inspected to remove rejection-quality images.\footnotemark 
\footnotetext[3]{Poor-quality images are usually attributed to  
image contamination by energetic electrons trapped in the earth's 
magnetosphere.}
The temperature determination uses the 171/195 filter ratio (Fig. 2a), so
a sequence of 171-195 image pairs was extracted from each data set
under the somewhat arbitrary constraint that each pair of images was
obtained no more than two minutes apart.  The resulting data set for
each loop typically contains about 50 image pairs.
The resulting intensity ratios were then coadded over the subimage
and over the time sequence to produce a single intensity ratio, with
associated error, for each data set.  The 171-\AA\ passband counts
were similarly coadded to produce a single count rate per pixel, with
associated error, for each data set.

Two types of errors result from this analysis.  First, there are 
errors due to noise in the data, which we consider to be 
based on Poisson statistics of data with
approximately 100 photons per data number.  Second, there is an error
associated with the width of the data distribution for each subimage; 
the distributions are approximately gaussian, and we take the 
corresponding error for each subimage to be one standard deviation of the data
distribution for that subimage.  For both the
filter ratio and passband count diagnostics, the noise error and
distribution error for each data set are on the order of
fractions of a percent; we thus consider the errors to be negligible
for the purposes of this study.

\section{Results and Discussion}

We first note some cautions/limitations regarding this analysis.
First, in using the filter ratios to determine temperatures, 
we implicitly assume that all the material
through which we look (i.e., integrated along the line of sight)
at each position along the loop is at the same density and temperature.
For this reason, we restrict ourselves to loop systems on the limb
and measure relatively isolated loops.  Second, analysis of the
density structure along loops is conceptually difficult because
of the intricate loop substructure evident in the images (Fig. 1).

The 171/195 filter ratios and 171-\AA\ count rates 
for each loop as a function of fractional
distance along the loop are given in Table 1.  As inspection of
Figure 2a indicates, the temperature as a function of the 171/195 
filter ratio is multivalued in the
coronal temperature range $\log(T)$ = $5 - 7$, so a definitive determination
of the loop temperature is not possible based on these data alone
(however, we note that the temperatures of maximum formation of 
the 171 and 195 lines are  $\log(T)$ =  6.0 and 6.2, respectively, so it is
tempting to conclude that the loop temperatures are around $\log(T) = 6.1$).
We note, however, that it is unlikely that the temperature profiles
change along the loops, since transition from, e.g., $\log(T) = 6.1$ to
$\log(T) = 6.4$ would result in observation of considerably lower 
intensity ratios at intermediate points along
the loop than are observed.  We thus conclude that there is no
significant temperature variation along the loops we consider.
Furthermore, Occam's razor suggests that all the
loops share the same temperature.

Theoretical loop models that include energy considerations predict a 
steep temperature rise in the transition
region and a small, but measurable, temperature rise
to a maximum at the loop top in the coronal part of the loop 
(see, e.g., Rosner et al. 1978; \cite{spv81}). 
In contrast, our observations show no
significant
temperature variation.  Figure 3 shows 
the temperatures and emission measures\footnotemark 
\footnotetext[4]{Observed emission measures are calculated using
\begin{equation}
     \mbox{EM} = \frac{\mbox{DN}/\mbox{s}/\mbox{pixel}}{\mbox{resp}_{171}(T)} \; \mbox{ cm}^{-5}  \ ,
\end{equation}  
where we use $\mbox{resp}_{171}[\log(T) = 6.1] = 2 \times 10^{-27}$ DN/s/pixel/EM (see Fig. 2b).
Theoretical emission measures are calculated using $\mbox{EM} = n_{e}^{2}D$, 
where $n_{e}$ is the electron number density and $D$ is the
line-of-sight depth.} for the observed 
loops and for three model loops: 
(1) an isothermal, hydrostatic loop with
$T = 1.34 \times 10^{6}$ K, $L = 10^{10}$ cm, 
a base emission measure of $6.25 \times 10^{27}$ cm$^{-5}$, and
uniform line-of-sight depth along the loop; (2) loop (1), but with a line-of-sight depth that
increases gradually along the loop by a factor of 4;  and 
(3) a static, steady-state, nonisothermal 
loop (cf. \cite{spv81}) that has $L = 10^{10}$ cm, base pressure chosen 
such that the loop-top temperature agrees with that of the observed 
loops, and uniform line-of-sight depth of $10^{10}$ cm; for model loop (3), 
the base proton number density is $2.4 \times 10^{10}$ cm$^{-3}$ and 
the base temperature is $2 \times 10^{4}$ K.

Figure 3a shows the near-constant observed loop temperatures 
and the slight rise in the temperature structure of model loop (3).
Figure 3b indicates that the observed emission-measure structure agrees 
better in its shape with the 
nonisothermal model (3) and in its magnitude with the isothermal 
models (1) and (2).
If the observations  
accurately reflect the temperature and emission-measure structure in the loop, 
it may be that physical process(es) 
not included in our
assumptions and model calculations exist in the
observed loops.  For example, the calculation for model loop (3) assumes a uniform 
volumetric heating rate, which may not describe actual loop heating. 
Furthermore, flows may introduce denser material
into the loops, and mixing may homogenize the overall structure; 
alternatively, the hydrostatic pressure balance may be strongly affected
by wave interactions with the background fluid (\cite{lr98}).

The energetic requirements of the loops we have
examined may range from $\sim 10^5 - 5 \times 10^6$ erg s$^{-1}$ cm$^{-2}$,
corresponding to line-of-sight depths of $10^{10} - 10^9$ cm; 
``standard'' values quoted in the
literature are typically $\lesssim 10^7$ erg s$^{-1}$ cm$^{-2}$ (cf. 
\cite{wn77}; \cite{vr78}).   Smaller line-of-sight depths may be
possible if the filling factor is small, as in the case of
filamentary emission;  such a configuration would correspond to
a higher, localized (filamentary) energy input at the base, 
consistent with localized 
heating events such as microflares.

We do not measure any
filter ratios consistent with transition-region temperatures of
$\sim 10^{4} - 10^{5}$ K; hence we conclude that the ``footpoint''
regions we choose lie above the transition region, a reasonable
conclusion given that the transition region occupies roughly 3 pixels 
per image, and is likely to be obscured by absorbing material along
the line of sight (Daw, DeLuca, \& Golub 1995).

An earlier study of loop temperature distributions using Yohkoh X-ray
data (\cite{kt96}) reports loop temperatures that increase from
the footpoints to maxima at the loop tops by factors of $\gtrsim 1.2$.
The loop temperature profiles we find vary by factors of at most 1.05.
The temperatures they measured are higher by a factor of $3 - 5$ than
the temperatures we report here.
The lack of temperature variation in the EUV loops considered here (see also 
\cite{gj75}; Aschwanden et al. 1999a, 1999b) 
invites speculation that there is a class of such isothermal
loops distinct from loops with a temperature maximum at the
apex.  Whether the difference is due to some fundamental physical
difference among loops, to a difference in the X-ray and EUV properties
of loops, or to some other effect warrants further investigation.

\acknowledgments

The authors thank Daniel Brown, Vinay Kashyap, Rebecca McMullen, and 
Clare Parnell 
for assistance and helpful discussions.  The paper benefited from helpful 
comments by the referee, Carole Jordan.  This work was supported
by a TRACE subgrant from Lockheed Martin to the University of Chicago
 and by Contract NAS5-38099 from NASA to LMATC.

\newpage

\begin{deluxetable}{cccccc}

\tablecaption{Loops Selected for Study}

\tablehead{ \colhead{Loop Letter} & \colhead{Date/Time}
        & \multicolumn{4}{c}{171/195 filter ratio} \nl 
  \colhead{(cf. Fig. 1)} & \colhead{} & 
     \multicolumn{4}{c}{and 171-\AA\ count rate (DN/s/pixel)} \nl
    \colhead{} & \colhead{} &
   \multicolumn{4}{c}{at fractional distance along loop} \nl
     \colhead{} & \colhead{} & \colhead{0.2 ( $\sim$ base)} & \colhead{0.3}
        & \colhead{0.7} & \colhead{1.0 (apex)}}

\startdata

a & 04 Jul 98  1800-2000h & 1.03 & 0.88 & 0.75 & 0.81 \nl
  &                       & 10.01 & 6.22 & 3.92 & 3.12 \nl
b & 26 Jul 98 2200-2300h  & 0.85 & 0.70 & 0.85 & 0.78  \nl
  &                       & 4.68 & 2.87 & 2.51 & 1.93 \nl
c & 18 Aug 98 1000-1100h  & 0.83 & 0.90 & 0.96 & 1.09  \nl
  &                       & 11.39 & 9.74 & 9.32 & 9.18 \nl
d & 20 Aug 98 0800-0900h  & 0.90 & 0.89 & 0.84 & 0.86  \nl 
  &                       & 10.20 & 7.50 & 5.84 & 6.35 \nl

\enddata

\end{deluxetable}

\newpage

\begin{figure}
\epsscale{0.8}
\vspace{-1in}
\plotone{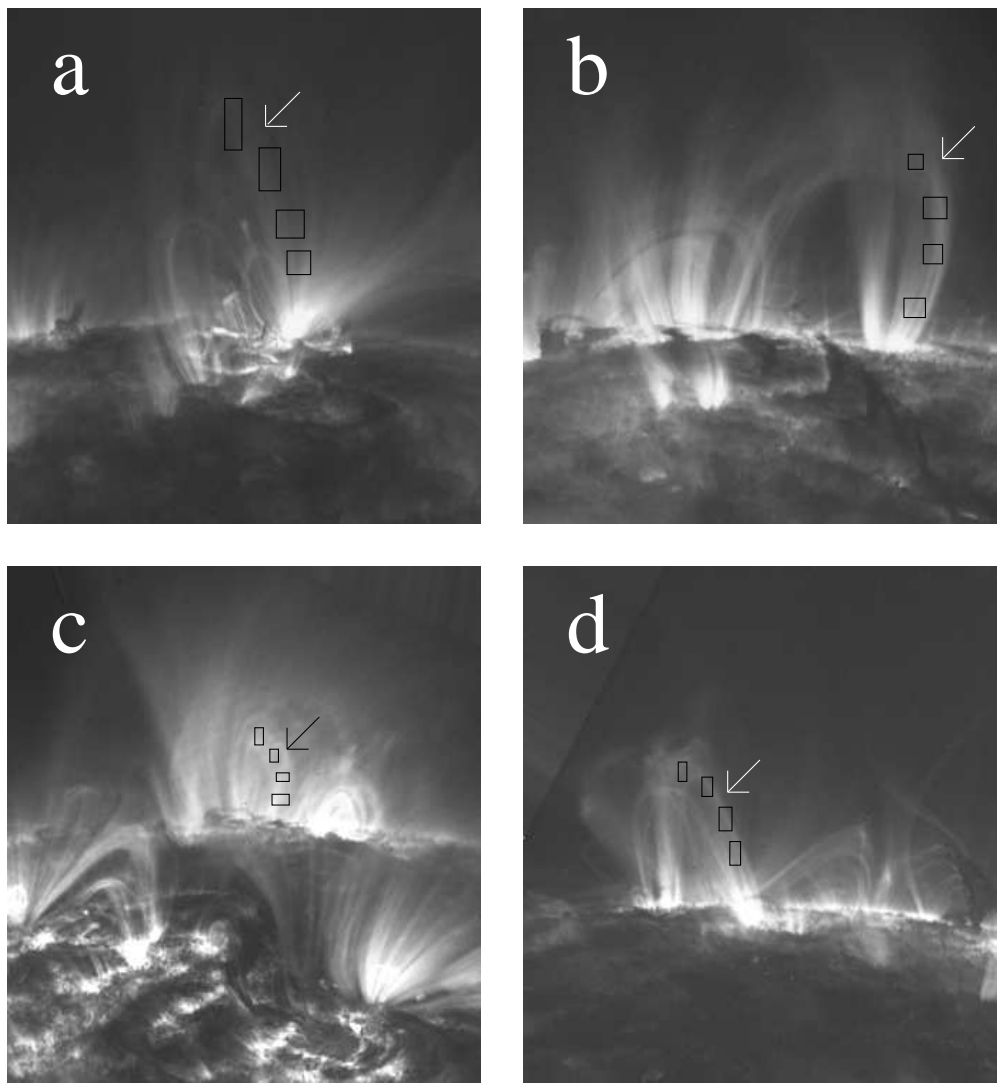}
\vspace{0.5in}
\caption{Images in the 171 \AA\ filter of the loops selected
     for this study (indicated by arrows): 
   (a) 04 Jul 1998, (b) 26 Jul 1998, (c) 18 Aug 1998,
        (d) 20 Aug 1998.  Images have been rotated so that the limb is
        roughly horizontal.  The field-of-view of each image is
        5\farcm6 $\times$ 5\farcm4 (roughly 680 $\times$ 650 pixels). 
        The rectangles represent the subimages used in this study.}
\end{figure}

\newpage

\begin{figure}
\epsscale{0.6}
\vspace{-0.5in}
\plotone{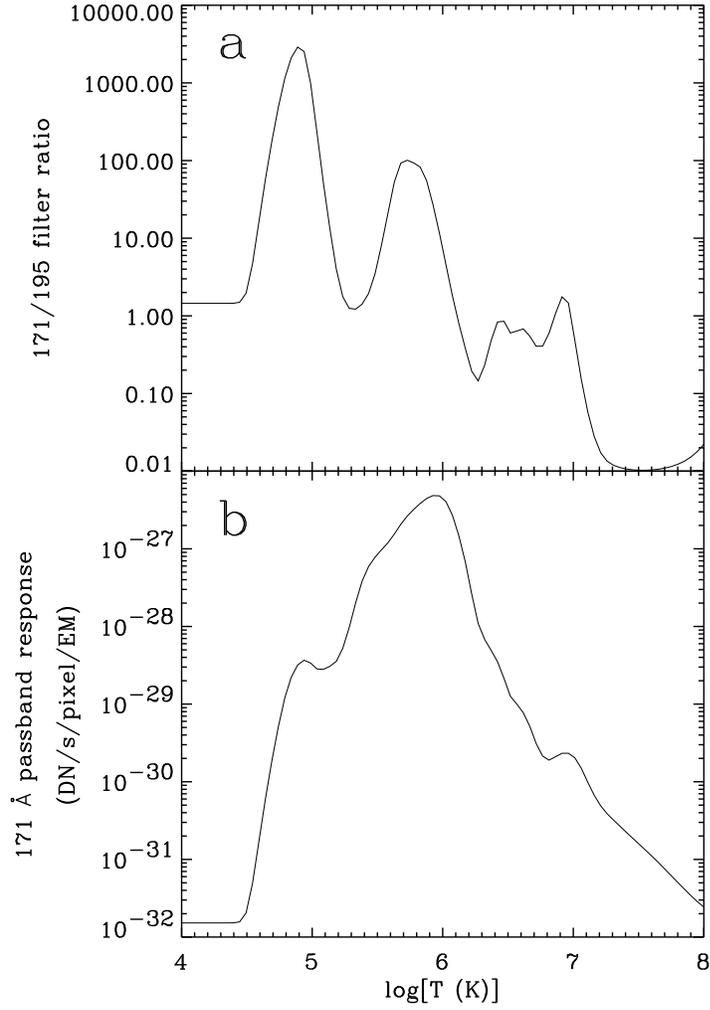}
\vspace{0.5in}
\caption{(a) Filter ratio of TRACE 171 \AA\ and 195 \AA\ passbands; (b) Response of 171 \AA\ passband in units of data number (DN) per second per pixel per emission measure (EM), where 1 DN $\approx 100$ photons and EM is
in units of cm$^{-5}$.}
\end{figure}

\newpage

\begin{figure}
\epsscale{0.6}
\vspace{-0.5in}
\plotone{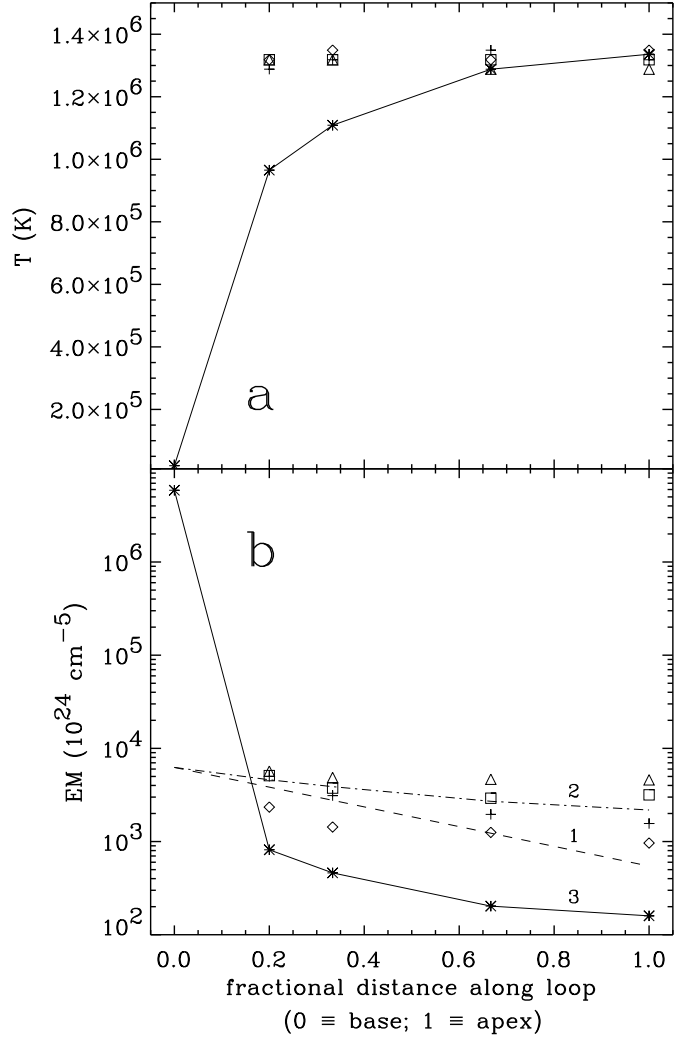}
\vspace{0.5in}
\caption{Temperature (a) and emission measure (b)
 as functions of fractional distance along the  
loop for loop (a) (plus signs), loop (b) (diamonds), loop (c) (triangles), 
loop (d) (boxes), and model loop (3) with $T$(apex) = $1.34 \times 10^{6}$ K and 
uniform line-of-sight depth $D = 10^{10}$ cm 
(connected asterisks).  Also shown in (b) are model loop (1): an 
isothermal ($T = 1.34 \times 10^{6}$ K), 
hydrostatic loop with a uniform line-of-sight depth (dashed line); and model loop (2): the same as model loop (1), but with a line-of-sight depth that increases along the
loop by a factor of 4 (dot-dashed line).  See \S 3 for discussion.  
The errors on the observations 
are comparable to or less than the
size of the plot symbols.}
\end{figure}

\end{document}